\documentclass[preprint,10pt]{elsarticle}
\usepackage{amssymb}
\journal{}
\usepackage{epsfig}
\usepackage{ifthen}
\usepackage{amsfonts}
\usepackage{amssymb}
\usepackage{amsmath}
\usepackage{amscd}
\usepackage{amsthm}
\usepackage{fancyvrb}
\usepackage{epsfig}
\usepackage{mathtools}
\usepackage{epsfig}
\usepackage{bbm}
\usepackage{fullpage}
\usepackage[boxed,algosection,linesnumbered]{algorithm2e}
\begin{document}
\begin{frontmatter}
%% Title, authors and addresses
%% use the tnoteref command within \title for footnotes;
%% use the tnotetext command for the associated footnote;
%% use the fnref command within \author or \address for footnotes;
%% use the fntext command for the associated footnote;
%% use the corref command within \author for corresponding author footnotes;
%% use the cortext command for the associated footnote;
%% use the ead command for the email address,
%% and the form \ead[url] for the home page:
%%
%% \title{Title\tnoteref{label1}}
%% \tnotetext[label1]{}
%% \author{Name\corref{cor1}\fnref{label2}}
%% \ead{email address}
%% \ead[url]{home page}
%% \fntext[label2]{}
%% \cortext[cor1]{}
%% \address{Address\fnref{label3}}
%% \fntext[label3]{}
\title{On Polynomial Time Absolute Approximation-bounded Solution of TSP and NP Complete Problems}
%% use optional labels to link authors explicitly to addresses:
\author[]{Wenhong Tian$^{a}$}
%\author[1]{GuoZhong Li}
%\author[1]{Xinyang Wang}
%\author[1]{Qin Xiong}
%\author[1]{YaQiu Jiang}
% \author[3]{Adel Nadjaran Toosi}
% \author[3]{Rajkumar Buyya}
\address[1]{School of Information and Software Engineering, \\University of Electronic Science and Technology of China (UESTC)}
%\address[2]{BigData Research Center of UESTC}
%\address[3]{CLOUDS Lab, Dept. of Computing and Information Systems, The University of %Melbourne, Australia}
%\author{}
%\address{}
\begin{abstract}
%% Text of abstract
The question of whether all problems in NP class are also in P class is generally considered one of the most important open questions in mathematics and theoretical computer science as it has far-reaching consequences to other problems in mathematics, computer science, biology, philosophy and cryptography. There are intensive research on proving `NP not equal to P' and `NP equals to P'. However, none of the `proved' results is commonly accepted by the research community up to date. In this paper, motived by approximability of traveling salesman problem (TSP) in polynomial time, we provide a polynomial time absolute approximation-bounded solution of TSP in Euclidean space. It may shine light on solving other NP complete problems in similar way.\\
\end{abstract}
\begin{keyword}
%% keywords here, in the form: keyword \sep keyword
NP problems\sep P Problems\sep  P versus NP\sep TSP\sep Polynomial Time Absolute Approximation Bounded Solutions
%% MSC codes here, in the form: \MSC code \sep code
%% or \MSC[2008] code \sep code (2000 is the default)
\end{keyword}
\end{frontmatter}
%%
%% Start line numbering here if you want
%%
% \linenumbers
% creates the second title. It will be ignored for other modes.
\section{Introduction}
P versus NP problem is one of seven Millennium Prize Problems in mathematics that were stated by the Clay Mathematics Institute [1] in 2000. As of Dec 2016, six of the problems remain unsolved. The official statement of P versus NP problem was given by Stephen Cook [2].  In computational complexity theory, Karp's 21 NP-complete problems are a set of computational problems which are NP-complete. In his 1972 paper [9], Richard Karp used Stephen Cook's 1971 theorem that the Boolean satisfiability problem is NP-complete (also called the Cook-Levin theorem) to show that there is a polynomial time many-one reduction from the Boolean satisfiability problem (BSP) to each of 21 combinatorial problems, thereby showing that they are all NP-complete. This was one of the first demonstrations that many natural computational problems occurring throughout computer science are computationally intractable, and it drove interest in the study of NP-completeness and the P versus NP problem. \\
%A correct solution to any of the problems results in a US 1,000,000 prize (sometimes called a Millennium Prize) being awarded by the institute [10]. 

Simply speaking, P problems mean that the class of problems can be solved exactly in polynomial time while NP (non-deterministic polynomial) problem stands for a class of problems which can not be solved in polynomial time. Intuitively, NP problem is the set of all decision problems for which the instances where the answer is ``yes" have efficiently verifiable proofs of the fact that the answer is indeed ``yes". More precisely, these proofs have to be verifiable in polynomial time by a deterministic Turing machine. In an equivalent formal definition, NP problems is the set of decision problems where the ``yes"-instances can be accepted in polynomial time by a non-deterministic Turing machine [18]. NP problems has far-reaching consequences to other problems in mathematics, biology, philosophy and cryptography.\\

The complexity class P is contained in NP, and NP contains many important problems. The hardest of which are NP-complete problems, whose solutions are sufficient to deal with any other NP problems in polynomial time. The most important open question in complexity theory, is the P versus NP problem which asks whether polynomial time algorithms actually exist for NP-complete problems and all NP problems. 
The important thing is that \textit{Karp showed that if any of them have efficient polynomial time algorithms, then they all do}. Many of these problems arise from real-world optimization problems including Sub Set Sum Problem (SSP), Traveling Salesman Problem (TSP), Bin Packing Problem (BPP), Hamiltonian Cycle Problem (HCP), and Chromatic Number Problem (CNP) etc.. Researchers later extend Karp's techniques to show hundreds, if not thousands of natural problems, are NP-complete.\\

It is widely believed that NP!=P in 2002 [4]. In 2012, 10 years later, the same poll was repeated [5]. The number of researchers who answered was 126 (83\%) believed the answer to be no, 12 (9\%) believed the answer is yes, 5 (3\%) believed the question may be independent of the currently accepted axioms and therefore is impossible to prove or disprove, 8 (5\%) said either don't know or don't care or don't want the answer to be yes nor the problem to be resolved. On the Web site \cite{gwoegiPvNP} , Prof. Gerhard Woeginger provides the unofficial archivist of about 116 claims for the NP vs P problem from 1986 to April 2016, among them, 49 (42\%) believed the answer to be no, 62 (53\%) believed the answer is yes, the other 5 (5\%) think Undecidable, or Unprovable or Unknow. About nine of papers in the list `established' NP=P by designing algorithms for variants of the TSP, though none of them is commonly accepted yet by the research community. \\

As for approximation of TSP, Christofides [Christofides,1976] provided an absolute 1.5-approximation algorithm. Arora [Arora, 1998] proposed an asymptotical (1+1/c)-approximation algorithm, but its computational complexity is of $O(n(logn)^{O(c)})$ where  $(logn)^{O(c)}$ can be huge since $c$ can be a few tens or larger, therefore it is not a practical algorithm but asymptotical bounded approximation (this is confirmed by Prof. Arora through email).  Tian et al. [Tian et al., 2016] introduced TGB algorithm with absolute approximation of (1+$\frac{1}{2}(\frac{\alpha+1}{\alpha+2})^{K-1}$)-approximation where $K$ is the number of iterations in TGB and $\alpha$ is the shape parameter of Generalized Beta distribution (introduce in Section 3) and can be obtained once a TSP is given.  In this paper, we focus on absolute approximation but not asymptotical approximation. For convenience, we just use approximation to represent absolute approximation.\\
In [Papadimitriou and Vempala, 2006], Papadimitriou and Vempala proved that, unless NP=P, there can be no polynomial-time $C$-approximation algorithm for the metric TSP with $C$ less than $\frac{220}{219}$, i.e., less than 0.45\%. In this paper,  we aim to propose is a absolute bounded approximation algorithm for TSP in Euclidean space. \\
% Most computer scientists seem to suspect P does not equal NP. 
% MIT computer scientist Scott Aaronson gives informal arguments against P=NP in his post, including his philosophical argument [ ].
%``If P=NP, then the world would be a profoundly different place than we usually assume it to be. There would be no special value in `creative leaps', no fundamental gap between solving a problem and recognizing the solution once it is found. Everyone who could appreciate a symphony would be Mozart; everyone who could follow a step-by-step argument would be Gauss; everyone who could recognize a good investment strategy would be Warren Buffett. It is possible to put the point in Darwinian terms: if this is the sort of universe we inhabited, why wouldn't we already have evolved to take advantage of it?" \\

%There are intensive research on proving `NP not equal to P' and `NP equals to P'. However, none of the `proved' results is commonly accepted by the research community yet up to %date. There do exist some problems previously believed to be in NP class but recently proved to be in P class, such as PRIME problem [20].
%graph isomorphism problem [ ]. 

The remaining sections are organized as follows. TSP is discussed in Section 2. Approximation bounded algorithm for TSP is proposed in Section 3. Our main results are provided in Section 4. Finally we conclude in Section 5.

\section{ TSP Formulation in Euclidean Space }

The TSP is one of most researched problems in combination optimization because of its importance in both academic need and real world applications. For surveys of the TSP and its applications, the reader is referred to [Cook, 2012] and references therein. 
We consider the $n$-node TSP defined in Euclidean space. This can be represented on a complete graph $G$= ($V,E$) where $V$ is the set of nodes and $E$ is the set of edges. The cost of an edge ($u$, $v$) is the Euclidean distance ($c_{uv}$) between $u$ and $v$. Let the edge cost matrix be $C[c_{ij}]$ which satisfies the triangle inequality. 

\textbf{Definition 1.} Symmetric TSP (STSP) is TSP in Euclidean distance (called ESTSP) and the edge cost matrix $C$ is symmetric.\\
\textbf{Definition 2.} Asymmetric TSP (ATSP) is TSP in Euclidean distance (called EATSP) and the edge cost matrix $C$ is asymmetric.

\textbf{Definition 3.} $\triangle$STSP is a STSP whose edge costs are non-negative and satisfies the triangle inequality, i.e., for any three distinct nodes (not necessary neighboring) ($i, j, k$), $(c_{ij}$+$c_{jk}) \geq c_{ik}$. The STSP is also called the \textbf{metric} TSP.

\textbf{Definition 4.} TSP tour. Given a graph $G$ in 2-dimensional Euclidean distance and its distance matrix $C$ where $c_{ij}$ denote the distance between node $i$ and $j$ (for both symmetric and asymmetric). A tour $T$ with $|V|$ nodes has length

\begin{equation}
L=\sum_{k=0}^{|V|-1} c_{T(k),T(k+1)}
\end{equation}

In 1977, Papadimitriou [Papadimitriou,1977] firstly proved that the Euclidean TSP (ETSP) is NP-complete by reduction of the Exact Cover Problem to the ETSP.

\section{On the Approximability of Metric TSP}
On the approximability of metric TSP, there is a well-known theorem as follow. \\
\textbf{Papadimitriou-Vempala Theorem}. In [Papadimitriou and Vempala,2006], Papadimitriou and Vempala (let us call it Papadimitriou-Vempala Theorem) proved that, unless NP=P, there can be no polynomial-time $C$-approximation algorithm for the metric TSP with $C$ less than $\frac{220}{219}$, i.e., less than 0.45\%. \\

%Similarly Sahni and Gonzalez [Sahni and Gonzalez, 1976] stated that if there is a polynomial time complexity heuristic %with an absolute performance ratio, then P=NP. \\

%However as we already see that LKH can solve all of TSPLIB instances to optimality in polynomial time. Even for the largest instances, 1,904,711-city problem for which optima are %not known yet, but the tour by LKH for 1,904,711-city was known to be no more than 0.0476\% longer than an optimal tour by Concorde bound [3]. This means that LKH solution is %below $\frac{220}{219}$ approximation. Therefore we think that Papadimitriou-Vempala Theorem needs reevaluation. 
Before continuing, the following two definitions are introduced:\\
\textbf{Definition 5.} maxTSP. The maximum tour length (B) is obtained using LKH where each edge cost ($c_{ij})$ is replaced by a very large value ($M$) minus the original edge cost, i.e., ($M$-$c_{ij}$). $M$ can be set as the maximum edge cost plus 1.\\
\textbf{Definition 6.} $k$-opt. is a local search with $k$-exchange neighborhoods and the most widely used heuristic method for the TSP.
$k$-opt is a tour improvement algorithm, where in each step $k$ links of the current tour are replaced by $k$ links in such a way that a shorter tour is achieved (see [Helsgaun 2009] for detailed introduction).\\
In the following, we  propose an algorithm called ITGBC which can obtain approximation ratio less than $\frac{220}{219}$ for metric TSP. 

We firstly propose a Generalized Beta (GB) distribution [Tian et al., 2016].
The probability density function (pdf) of GB is defined as
\begin{equation}
f(x,\alpha,\beta, A, B)=\frac{(x-A)^{\alpha-1}(B-x)^{\beta-1}}{Beta(\alpha,\beta)}\label{eq:GBDensity1}
\end{equation}
where $Beta(\alpha,\beta)$ is the beta function
\begin{equation}
Beta(\alpha,\beta)=\int_{0}^{1}t^{\alpha-1}(1-t)^{\beta-1}dt \label{eq:beta1},
\end{equation}
%$Beta(\alpha,\beta)$=$\int_{0}^{1}t^{\alpha-1}(1-t)^{\beta-1}dt$,
$A$ and $B$ is the lower bound and upper bound respectively, $\alpha>0$, $\beta>0$. For TSP, $A$ and $B$ represents the minimum and maximum tour length (maxTSP) respectively. \\

Some of the following results are introduced in [Tian et al, 2016], for completeness, we restate here and introduce \textbf{Iterative Truncated Generalized Beta distribution Based on Christofides Algorithm (ITGBC)} firstly. ITGBC algorithm performs in seven steps: \\

\begin{itemize}

\item

(1). Finding the minimum spanning tree $MST$ of the input graph $G$ representation of metric TSP;

\item

(2). Taking $G$ restricted to vertices of odd degrees in $MST$ as the subgraph $G^{*}$; This graph has an even number of nodes and is complete;

\item

(3). Finding a minimum weight matching $M^{*}$ on $G^{*}$;

\item

(4). Uniting the edges of $M^{*}$ with those of the $MST$ to create a graph $H$ with all vertices having even degrees;

\item

(5). Creating a Eulerian tour on $H$ and reduce it to a feasible solution using the triangle inequality, a short cut is a contraction of two edges ($i, j$) and ($j, k$) to a single edge ($i, k$); 
\item
(6). Applying Christofides algorithm [Christofides,1976] to a ESTSP forms a truncated GB (TGB) for the probability density function of optimal tour lengths, with expectation (average) value at most 1.5OPT-$\epsilon$, where $\epsilon$ is a very small value; Applying $k$-opt to the result of Christofides algorithm forms another TGB for probability density function of optimal tour lengths; 
\item
(7). Iteratively applying this approach, taking the expectation value of $(K-1)$-th iteration as the upper bound ($\hat{b_K}=\frac{\mu_t^{K-1}-A}{B-A}$) of the $K$-th iteration, we have the expectation value (denoted as $\mu_t^K$) after $K$ iterations ($K\geq 2$),  which is proved in [Tian et al., 2016],
\begin{align}
\mu_t^K&=A+(B-A) \frac{B_2(0,\hat{b_K},\alpha+1,\beta)}{B_2(0,\hat{b_K},\alpha,\beta)} \nonumber\\
&= A+(B-A)g(\hat{b_K})\nonumber\\
& \leq (1+\frac{1}{2}(\frac{\alpha+1}{\alpha+2})^{K-1})A
\end{align}
%\begin{equation}
%g(\hat{b_K})=\frac{B_2(0,\hat{b_K},\alpha+1,\beta)}{B_2(0,\hat{b_K},\alpha,\beta)} \leq \frac{\alpha+1}{\alpha+2} \hat{b_K}=\frac{0.5A(\frac{\alpha+1}{\alpha+2})^{K-1}}{B-A}, ~when~n>20~and~\hat{b_K}<0.5
%\end{equation}
\begin{equation}
B_2(0,t,\alpha,\beta)=\int_{0}^{t} x^{\alpha-1}(1-x)^{\beta-1} \mathrm{d}t \label{eq:TruncatedBy}
\end{equation}

\end{itemize}

\textbf{Theorem 1.} ITGBC algorithm is (1+$\frac{1}{2}(\frac{\alpha+1}{\alpha+2})^{K-1}$)-approximation where $K$ is the number of iterations in ITGBC algorithm, $\alpha$ is the shape parameter of TGB and can be determined once ETSP instance is given. \\
The present author proved Theorem 1 in [Tian et al., 2016] , for completeness, we provide the proof in the following.

\begin{figure} [htp!]

\begin{center}

%\hfill

{\includegraphics [width=0.7\textwidth,angle=-0] {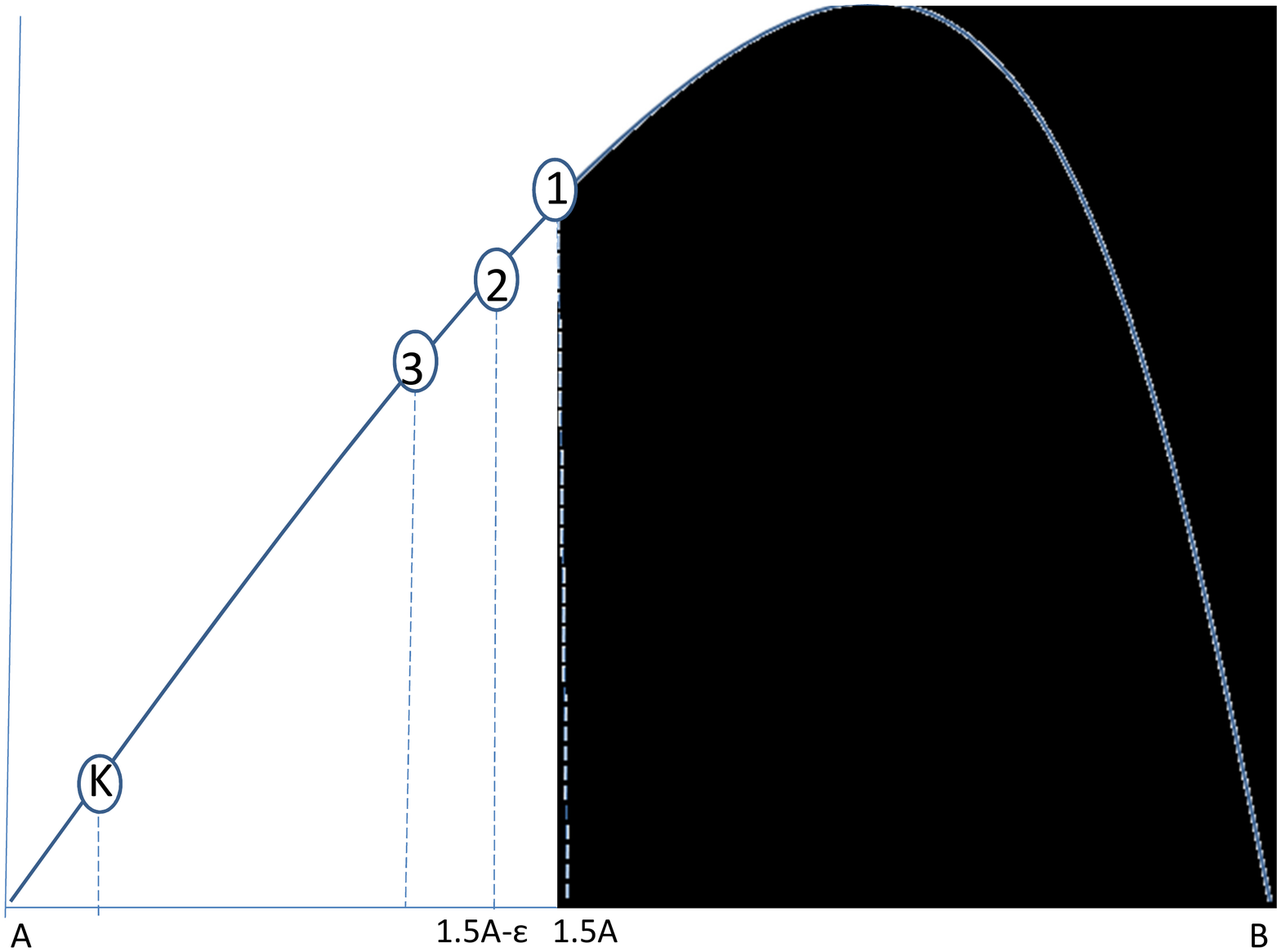}}

%\hspace*{\fill}

\caption{The   process in ITGBC algorithm }

\end{center}

\end{figure}
\begin{proof}

Applying $k$-opt to the result obtained by Christofide algorithm as shown in Fig.1. The TGB in this case is truncated from above. Denote the first truncation by Christofides' algorithm as the first truncation ($K$=1). The probability density function of the second TGB is given by
\begin{equation}
f_t^2(x,\alpha,\beta, A, B,a_2,b_2)=\frac{(x-A)^{\alpha-1}(B-x)^{\beta-1}} {\int_{a_2}^{b_2} (x-A)^{\alpha-1}(B-x)^{\beta-1}}\label{eq:TruncatedBetaDensity}
\end{equation}
In this case, $a_2$=$A$, $b_2$=1.5$A$ because the distribution is based on the result after applying Christofides algorithm which assures the upper bound is at most 1.5$A$, see Fig.1.
Setting $\hat{x}$=$\frac{x-A}{B-A}$, $\hat{a_2}$=$\frac{A-A}{B-A}$=0, $\hat{b_2}$=$\frac{1.5A-A}{B-A}$=$\frac{0.5A}{B-A}$, we have
\begin{align}
C_0 &= {\int_{a_2}^{b_2} (x-A)^{\alpha-1}(B-x)^{\beta-1}}\mathrm{d}x \nonumber\\
&={\int_{0}^{\hat{b_2}} ((B-A)\hat{x})^{\alpha-1}((B-A)(1-\hat{x})^{\beta-1}}\mathrm{d}x \nonumber\\
&=(B-A)^{\alpha+\beta-1}B_2(0,\hat{b_2},\alpha,\beta)
\end{align}

where
\begin{equation}
B_2(0,t,\alpha,\beta)=\int_{0}^{t} x^{\alpha-1}(1-x)^{\beta-1} \mathrm{d}t \label{eq:TruncatedBy}
\end{equation}
By the definition of the expectation or mean value (denoted as $\mu_t^2$) for $f_t^2(x,\alpha,\beta, A, B,a_2,b_2)$, we have
\begin{align}
\mu_t^{2}-A=\int_{a_2}^{b_2}(x-A)f_t^2(x,\alpha,\beta, A, B,a_2,b_2)\mathrm{d}x \nonumber\\
= \frac{ {\int_{a_2}^{b_2} (x-A)^{\alpha}(B-x)^{\beta-1}}\mathrm{d}x}{C_0} \nonumber\\
=  \frac{(B-A)^{\alpha+\beta}B_2(0,\hat{b_2},\alpha+1,\beta)}{C_0} \nonumber\\
= (B-A)\frac{B_2(0,\hat{b_2},\alpha+1,\beta)}{B_2(0,\hat{b_2},\alpha,\beta)} \nonumber\\
=>\mu_t^2=A+(B-A)\frac{B_2(0,\hat{b_2},\alpha+1,\beta)}{B_2(0,\hat{b_2},\alpha,\beta)}\label{eq:mu2}
\end{align}

Taking the expectation value of $(K-1)$-th iteration as the upper bound ($\hat{b_K}=\frac{\mu_t^{K-1}-A}{B-A}$) of the $K$-th iteration, we apply this approach iteratively and have the expectation value after $K$ iterations ($K\geq 2$), denoted as $\mu_t^K$,
\begin{align}
\mu_t^K&=A+(B-A) \frac{B_2(0,\hat{b_K},\alpha+1,\beta)}{B_2(0,\hat{b_K},\alpha,\beta)} \nonumber \\
&= A+(B-A)g(\hat{b_K})\nonumber
% & \leq (1+\frac{1}{2}(1-\frac{1}{\alpha+2})^k)A
\end{align}
Notice that the expectation value of the ($K$-1)-iteration is taken as the upper bound ($\hat{b_K}=\frac{\mu_t^{K-1}-A}{B-A}$ here $A$ is OPT and $B$ is the maxTSP) of the $K$-iteration, as shown in Fig.1.
Setting
%\begin{equation}
%\hat{b_{k}}=\frac{1.5A-k\epsilon-A}{B-A}=\frac{0.5A(1-\frac{1}{\alpha+2})^k}{B-A}
%\end{equation}

\begin{equation}
g(\hat{b_K})=\frac{B_2(0,\hat{b_K},\alpha+1,\beta)}{B_2(0,\hat{b_K},\alpha,\beta)}
% \leq \frac{\alpha+1}{\alpha+2}\hat{b_k}=(1-\frac{1}{\alpha+2}) \hat{b_k}
\end{equation}
The exact expression of $g(\hat{b_K})$ can be stated in a hypergeometric series, and

\begin{equation}
B_2(0,\hat{b_K},\alpha,\beta)=\frac{\hat{b_K}^\alpha}{\alpha}F(\alpha,1-\beta,\alpha+1,\hat{b_K})
\end{equation}
and $ F(a,b,c,x)$

\begin{align}
&=1+\frac{ab}{c}x+\frac{a(a+1)b(b+1)}{c(c+1)2!}x^2 \nonumber \\
& +\frac{a(a+1)(a+2)b(b+1)(b+2)}{c(c+1)(c+2)3!}x^3+...
%&\approx \frac{ab}{c}x, ~when~x<0.5, ab/c>100
\end{align}
%When $x$ is small (less than 0.5 for all cases) and $\alpha$, $\beta$ is large (from a few tens to a few %hundreds in all cases), we can take the second term in above equation to approximate $F(a,b,c,x)$, and %set $\hat{b_K}$=$\frac{u_t^{K-1}-A}{B-A}$ for $K\geq 2$, this is because that the expectation value of %$(K-1)$-th iteration is taken as the upper bound of $K$-th truncation. We find that
In all cases, we have $\alpha>$1, $\beta>$1, $\hat{b_K}\in (0,1)$ as shown in [Tian et al., 2016], therefore $ F(a,b,c,x)$ is an monotonic decreasing function. We have
%\begin{equation}
%g(\hat{b_K})\leq \frac{\alpha+1}{\alpha+2} \hat{b_K}
%\end{equation}
\begin{equation}
u_t^2= A+(B-A)g(\hat{b_2})\leq A+0.5A \frac{\alpha+1}{\alpha+2}
\end{equation}

continue this for $g(\hat{b_3})$, $u_t^3$, $g(\hat{b_4})$, $u_t^4$,..., so forth, we have
\begin{equation}
\hat{b_K}\leq \frac{0.5A}{B-A} (\frac{\alpha+1}{\alpha+2} )^{K-1}
\end{equation}
and 
\begin{align}
g(\hat{b_K}) &=\frac{B_2(0,\hat{b_K},\alpha+1,\beta)}{B_2(0,\hat{b_K},\alpha,\beta)} \nonumber \\
& \leq \frac{\alpha+1}{\alpha+2} \hat{b_K} \nonumber \\
& =\frac{0.5A(\frac{\alpha+1}{\alpha+2})^{K-1}}{B-A}, %~when~n>20~and~\hat{b_K}<0.5
\end{align}
Therefore
\begin{align}
\mu_t^K&=A+(B-A) \frac{B_2(0,\hat{b_K},\alpha+1,\beta)}{B_2(0,\hat{b_K},\alpha,\beta)} \nonumber\\
&= A+(B-A)g(\hat{b_K})\nonumber\\
& \leq (1+\frac{1}{2}(\frac{\alpha+1}{\alpha+2})^{K-1})A % ~when~n>20~and~\hat{b_K}<0.5
\end{align}
This completes the proof.
\end{proof}

%One can see that as $K$ increases, the approximation ratio (1+$\frac{1}{2}(\frac{\alpha+1}{\alpha+2})^{K-1}$) can be %less than $\frac{220}{219}$. Actually when $K>$ (1+$\frac{log0.009}{log(1-1/(\alpha+2))}$), the approximation ratio will be less than %$\frac{220}{219}$. \\

\textbf{ Theorem 2. The computational complexity of ITGBC algorithm is of \\ $O(max (n^3, K(k^3+k\sqrt{n})))$. } \\ 
\textbf{Proof:} In [Helsgaun, 2009], a method with computational complexity of $O(k^3+k\sqrt{n})$ is introduced for $k$-opt. Since ITGBC applies Christofides algorithm firstly which has computational complexity of $O(n^3)$ [Christofides,1976], and then applies $k$-opt with $K$ iterations in LKH which has computational complexity of $O(K(k^3+k\sqrt{n}))$, the computational complexity of LKH is estimated to be $O(n^{2.2})$ [Helsgaun, 2009], so altogether the computational complexity of ITGBC is of $O(max (n^3, K(k^3+k\sqrt{n})))$. \\

\subsection{Our Main Results}
%\textbf{Observation 1.} For all known metric TSPs, there exist polynomial time algorithms such as LKH and ITGBC to %solve them in polynomial time with approximation ratios less than $\frac{220}{219}$.\\

\textbf{Theorem 3.} Metric TSP is one of NP complete problem [Papadimitriou, 1977], which can be solved by  ITGBC in polynomial time $C$-approximation with $C$ less than $\frac{220}{219}$.  \\
\begin{proof}
From Theorem 1, it can been seen the approximation ratio (1+$\frac{1}{2}(\frac{\alpha+1}{\alpha+2})^{K-1}$) can be less than $\frac{220}{219}$. Actually when the instance is given, the shape parameter $\alpha$ can be estimated easily as shown in [Tian et al., 2016]. By fixing the approximation ratio to obtain the number of iterations in ITGBC, through a simple numerical computation, we know when $K>$ (1+$\frac{log0.009}{log(1-1/(\alpha+2))}$), the approximation ratio will be less than $\frac{220}{219}$. \\
\end{proof}
According to Papadimitriou-Vempala Theorem, this happens only when NP=P. This may imply  NP=P.

\section {Numerical Results}
For implementation, ITGBC algorithm is based on Christofides' algorithm and LKH  source codes, so it takes both advantages of them and provides approximation bounded results. 
\subsection{Polynomial Time Approximation-Bounded Solutions to ESTSP}
The results in Table 1 are obtained from William J. Cook's book [Cook, 2012], Chapter 1, where all problems are solved to optimality by different tools except for 1000,000 city problem and 1,904,711 city problem, for which optima are not known yet.\\
\begin{table}[ht]
\caption{TSP Records Variation By Years [Cook, 2012]} % title of Table
\centering % used for centering table
\begin{tabular}{c c c c} % centered columns (4 columns)
\hline\hline %inserts double horizontal lines
\# Nodes & Year (Solved) &Description & Authors \\[1ex]
%heading
\hline % inserts single horizontal line
48 & 1954& USA cities& Dantzig et al.(by hand) \\
64 & 1971& random nodes & Micheal Held, Richard Karp\\
80 & 1975& random nodes & Panagiotis Miliotis\\
120& 1977& Germany cities& Martin Grotschel, Manfred Padberg\\
318& 1987& cities & Manfred Padberg,Harlan Crowder\\
532& 1987& USA cities& Martin Grotschel, Manfred Padberg\\
666& 1987& World cities & Martin Grotschel, Manfred Padberg\\
1002& 1987& cities & Martin Grotschel, Manfred Padberg\\
2392& 1987& cities& Martin Grotschel, Manfred Padberg\\
%2600& 1988& TSP race& \\ 
3038 & 1992& cities & Concorde \\
13509& 1998& USA cities& Concorde\\
15112& 2001& cities& Concorde\\
24978& 2004& Sweden cities& Concorde\\
85900& 2006& cities& Concorde, LKH [Helsgaun,2009]\\
100000& 2009* & Japan & Yuchi Nagata\\
1904711& 2010*& World TSP Challenge& LKH [Helsgaun,2009] \\[1ex]

\hline %inserts single line
\end{tabular}
\label{table:nonlin} % is used to refer this table in the text
\end{table}
In Table 1, Nagata's tour for 1000,00-city Mona Lisa tour is known to be at most 0.0003\% longer than an optimal solution; The tour by LKH [Helsgaun,2009] for 1,904,711-city of length 7,515,790,345 meters was known to be no more than 0.0476\% longer than an optimal tour. \\

\textbf{Definition 7. Concorde Algorithm [Concorde,2003]:} {Concorde is a computer code for the STSP and some related network optimization problems. The code is written in the ANSI C programming language. Concorde's TSP solver has been used to obtain the optimal solutions to the full set of 110 TSPLIB instances, the largest having 85,900 cities. Executable versions of Concorde with qsopt for Linux and Solaris are available [Concorde,2003]. Hans Mittelmann has created a NEOS Server (http:\//neos-server.org\/) for Concorde, allowing users to solve TSP instances online.} \\
\textbf{Definition 8. LKH algorithm [Helsgaun,2009]}: {LKH is an effective implementation of the Lin-Kernighan heuristic [Lin and Kernighan,1973] for solving the traveling salesman problem.
Computational experiments have shown that LKH is highly effective. LKH has produced optimal solutions for all solved problems they have been able to obtain; including a 85,900-city instance (at the time of writing, the largest nontrivial instance solved to optimality). Furthermore, the algorithm has improved the best known solutions for a series of large-scale instances with unknown optima, among these a 1,904,711-city instance (called World TSP).} \\
Both Concorde [Concorde, 2003] and LKH [Helsgaun,2009] solve all 110 TSPLIB instances [Reinelt,1991] to optimums. \\
%As for Concorde, till 2003 it produced optimal solutions for 106 of TSPLIB instances, except for four instances brd14051, d18512,pla33810 and pla85900.

\begin{table}[ht]
\caption{5 Longest Running Time TSPLIB Instances Solved Exactly by LKH [Helsgaun,2009]} % title of Table
\centering % used for centering table
\begin{tabular}{c c c } % centered columns (4 columns)
\hline\hline %inserts double horizontal lines
Name & \#Nodes &Running Time (Seconds) \\[1ex]
%heading
\hline % inserts single horizontal line
fl1577& 1577& � 10975�\\
fnl4461& 4661& 10973 \\
u1817 & 1817 & 2529�\\
pcb3038& 3038& 3237� \\
pla7397 & 7397& 130220� \\[1ex]
\hline %inserts single line
\end{tabular}
\label{table:nonlin} % is used to refer this table in the text
\end{table}

Table 2 shows that LKH results for 5 STSP TSPLIB instances which are top 5 longest running time instances for LKH solved in 1998 and running times are measured in seconds on a 300 MHz G3 Power Macintosh.

%Table 3 shows that Concorde results for the same TSPLIB instances, where the tests were carried out on a 500 MHz Compaq XP1000 workstation in 2003.

%\begin{table}[ht]
%\caption{5 STSP TSPLIB instances solved by Concorde [4]} % title of Table
%\centering % used for centering table
%\begin{tabular}{c c c } % centered columns (4 columns)
%\hline\hline %inserts double horizontal lines
%Name & \#Nodes &Running Time (Seconds) \\[1ex]
%%heading
%\hline % inserts single horizontal line
%fl1577& 1577& � 6705.04�\\
%fnl4461& 4661& 53420.13� \\
%u1817 & 1817 & 449230.55�\\
%pcb3038& 3038& 80828.87� \\
%pla7397 & 7397& 428996.2� \\[1ex]
%\hline %inserts single line
%\end{tabular}
%\label{table:nonlin} % is used to refer this table in the text
%\end{table}

\section{Conclusions and Future Work}

%One can see from Table 1 that, the scale (the number of nodes) of TSP is increased as year increasing; the TSP becomes harder because of the scale becomes %larger and larger. This raises a question that what is good bound (the largest number of nodes) for NP=P in case of TSP. Since as $n$ (the number of nodes) %becomes very large, even the polynomial time computational complexity problem will become very difficult to solve. \\
In this paper, we provide an algorithm ITGBC in polynomial time with absolute approximation bounded solutions for TSP. One can see from Table 1 that, the scale (the number of nodes) of TSP is increased as year increasing; one of reasons for the TSP to become harder is because of the scale becomes larger and larger. For TSPLIB instances with node number less than 5000, ITGBC based on LKH can solve them to optimality in less than a few hours or shorter. These mean that ITGBC can provide exact or approximation-bounded solutions to practical TSPs. \\
How about other NP problems? Can they also be solved in similar way? According to Karp's result [Karp,1972] that if any of NP problems have efficient algorithms, then they all do. Hopefully our proposed approach can shine light on other NP problems. 

% use section* for acknowledgement
\section*{Acknowledgments}
This research is partially supported by China National Science Foundation (CNSF) with project ID 61672136, 61650110513; and Xi Bu Zhi Guang Project (R51A150Z10). 
A version of manuscript is posted on arxiv at https://arxiv.org/abs/1605.06183.
\bibliographystyle{elsarticle-num}
%\bibliography{<your-bib-database>}

\section*{References}
\begin{thebibliography}{00}
%% \bibitem must have the following form:
%% \bibitem{key}...
%%
\bibitem{Claymath2000}
Clay Mathematics Institute, http://www.claymath.org/millennium-problems/millennium-prize-problems.

\bibitem{Cook1971}
Stephen Cook (1971). \emph{The Complexity of Theorem Proving Procedures}, Proceedings of the third annual ACM symposium on Theory of computing. pp. 151–158, March of 1971.
\bibitem{Cook2012}
William Cook, In Pursuit of the Traveling Salesman, Princeton University Press, 2012.
\bibitem{Concorde}
Concorde, http://www.math.uwaterloo.ca/tsp/concorde.html
%\bibitem{Christofides1976}
% N. Christofides, Worst-case analysis of a new heuristic for the travelling salesman problem, Report 388, Graduate School of Industrial Administration, CMU, 1976.
\bibitem{Gasarch2002}
William I. Gasarch, (June 2002). \emph{The P=?NP poll} , SIGACT News 33 (2): 34-47. doi:10.1145/1052796.1052804. Retrieved 2008-12-29.
\bibitem{Gasarch2012}
William I. Gasarch, \emph{The Second P=?NP poll}, SIGACT News 74, 2012.
%\bibitem{Grahm1969}
%R. L. Graham, \emph{Bounds on Multiprocessing Timing Anomalies}, SIAM Journal on Applied Mathematics, Vol.17, %No.2, pp.416-429, 1969.
\bibitem{Helsgaun2009}
K. Helsgaun,General k-opt submoves for the Lin-Kernighan TSP heuristic. Mathematical Programming Computation, 2009,doi: 10.1007/s12532-009-0004-6.
\bibitem{Karp1972}
Richard M. Karp (1972), \emph{Reducibility Among Combinatorial Problems}, In R. E. Miller and J. W. Thatcher (editors). Complexity of Computer Computations. New York: Plenum. pp. 85-103.

\bibitem{Lin1973}
S. Lin, B.W. Kernighan, An effective heuristic algorithm for the traveling-salesman problem. Oper. Res.21, 498-516 (1973).
\bibitem{LKHCode}
LKH codes, http://www.akira.ruc.dk/~keld/research/LKH/, last accessed Jan. 15th of 2015.

\bibitem{Papadimitriou1977}
Christos H. Papadimitriou, The Euclidean travelling salesman problem is NP-complete, Theoretical Computer Science, Volume 4, Issue 3, June 1977, Pages 237-244.
\bibitem{Papadimitriou2006}
Christos H. Papadimitriou and Santosh Vempala, On the approximability of the traveling salesman problem,Combinatorica 26 (1) (2006) 101-120.
\bibitem{Reinelt1991}
G. Reinelt. TSPLIB, a traveling salesman problem library. ORSA Journal on Computing, 3(4):376-384, 1991.
\bibitem{Reiter1966}
S. Reiter and D. B. Rice. Discrete optimizing solution procedures for linear and nonlinear integer programming problems. Management Science, 12(11):829-850, 1966.
\bibitem{Role2008}
David Soler, A transformation for the mixed general routing problem with turn penalties,Journal of the Operational Research Society 59, 540-547, April 2008.
\bibitem{Tian2015}
Wenhong Tian, Chaojie Huang, Xinyang Wang, A Near Optimal Approach for Symmetric Traveling Salesman Problem in Eclidean Space, To appear in Proceedings of ICORES 2017, Portugal, also available at  arxiv https://arxiv.org/pdf/1502.00447.pdf
\bibitem{wiki}
Wikipedia, http://en.wikipedia.org/wiki/NP-complete.
\bibitem{gwoegiPvNP}
http://www.win.tue.nl/~gwoegi/P-versus-NP.htm
\bibitem{Neeraj2002}
Manindra Agrawal, Neeraj Kayal, Nitin Saxena, PRIMES is in P, IIT Kanpur, Preprint of August 8, 2002, http://www.cse.iitk.ac.in/news/ primality.html.
\bibitem{Christofides1976}
N. Christofides, Worst-case analysis of a new heuristic for the travelling salesman problem, Report 388, Graduate School of Industrial Administration, CMU, 1976.
%\bibitem{Sahni1976}
%Sartaj Sahni and Teofilo Gonzalez. P-complete approximation problems. Journal of the ACM, 23(3):555-565, July 1976.
\bibitem{Arora1998}
Sanjeev Arora, Polynomial Time Approximation Schemes for Euclidean Traveling Salesman and other Geometric Problems. J. ACM 45(5):753-782,1998.
\end{thebibliography}
%% Authors are advised to submit their bibtex database files. They are
%% requested to list a bibtex style file in the manuscript if they do
%% not want to use elsarticle-num.bst.
%% References without bibTeX database:

\end{document}